\documentclass[12pt]{article}

\usepackage{amsmath}
\usepackage{amssymb}

\usepackage{graphicx}

\usepackage{cite}

\usepackage{color} 

\usepackage[labelfont=bf,labelsep=period,justification=raggedright]{caption}

\makeatletter
\renewcommand{\@biblabel}[1]{\quad#1.}
\makeatother

\date{}

\begin{document}

\begin{flushleft}
{\Large
\textbf{A network model of cellular aging}
}
\\
Hong Qin$^{1, \ast}$ 
\\
\bf{1} Biology Department, Spelman College, Atlanta, GA 30314, U.S.A.
\\
$\ast$ E-mail: hqin@spelman.edu
\end{flushleft}

\section*{Abstract}
What is aging? Mechanistic answers to this question remain elusive despite decades of research. Here, we propose a mathematical model of cellular aging based on a model gene interaction network. Our network model is made of only non-aging components - the biological activities of gene interactions decrease with a constant mortality rate. Death of a cell occurs in the model when an essential gene loses all of its interactions to other genes, equivalent to the deletion of an essential gene. Gene interactions are stochastic based on a binomial distribution. We show that the defining characteristic of biological aging, the exponential increase of mortality rate over time, can arise from this gene network model during the early stage of aging.  Hence, we demonstrate that cellular aging is an emergent property of this model network. Our model predicts that the rate of aging, defined by the Gompertz coefficient, is approximately proportional to the average number of active interactions per gene and that the stochastic heterogeneity of gene interactions is an important factor in the dynamics of the aging process. This theoretic framework offers a mechanistic foundation for the pleiotropic nature of aging and can provide insights on cellular aging.


\section*{Introduction}

Aging is a fundamental question in biology \cite{Williams57, Finch90, Kirkwood00Nature, PatridgeBarton93Nature}, and is observed in \textit{E. coli} \cite{Wang10CB, StewartTaddei05, Nystrom07Plos},  the budding yeast \cite{Mortimer59}, worms \cite {Kenyon1996Ponce}, fruit flies \cite{Helfand03fly}, mice and humans \cite {Finch90,PatridgeBarton93Nature, Kim07aging}.
Aging also occurs in complex machinery, such air planes and auto mobiles. 
In general, aging can be quantitatively defined by mortality rate $\mu (t) $, which is  the normalized declining rate of viability $S(t)$,
\begin{align}\label{Eq One}
  &\text {Mortality rate: } \quad & & \mu(t) = - \frac{1}{S(t)} \frac{dS(t)}{dt},  &
\end{align}
where $t$ is time. 
Mortality rate $\mu(t)$ describes the chance of dying over age, and aging occurs when mortality rate is a positive increasing function of time.
Mortality rate is also known as the force of mortality, failure rate, hazard rate, and intensity function in various contexts \cite{Gumbel04, BarlowProschan96,  Leemis09}. 
Viability $S(t)$ is also known as the survival function and can be found from the mortality rate as $S(t) =  e^{ \int_{t=0}^{t} -\mu(t)dt} $. 
From viability, we can find the cumulative distribution function of lifespan as $1-S(t)$. 

Mortality rate $\mu(t)$ is often an exponential function of time for biological aging, known as the Gompertz model \cite{Gompertz1825, Olshansky1997Demography},
and a power function of time for machine aging, known as the Weibull model \cite{Weibull1951}: 
\begin{align} 
  &\text{Gompertz model:} & &\mu (t) =  R e^{Gt}     ,  & \label {Eq Gompertz} \\
  &\text{Weibull model:}    & &\mu (t) =  c_1 t^{c_2}  .  & \label {Eq Weibull}  
\end{align}
In the Gompertz model, $R$ is the initial mortality rate when $t$ is zero, and $G$ is the Gompertz coefficient. 
The initial mortality rate $R$ can be interpreted as the lifespan potential at birth. 
The Gompertz coefficient $G$ has a unit of 1/time, describes the acceleration of mortality rate $\mu$ over time, and hence is a measure for the rate of aging.  
In the Weibull model, $c_1$ and $c_2$ are constants. 

Using a classical reliability model of a serial circuit, Gavrilov and Gavrilova showed that the Weibull model of aging occurs in relatively homogeneous systems such as machinery, and the Gompertz model of aging occurs in relatively heterogeneous systems such as organisms in biology \cite{GG01} (This work will be referred to as GG01 hereafter). 
Other forms of mortality rate functions, such as the logistic function \cite{Vaupel98Sci}, were often used, especially when the studies focus on late life during aging.  
We will focus on the Gompertz and  Weibull models because we are interested in the emergence of aging during the early life stage and because of their theoretic implications from the reliability perspective.  
Moreover, the Gompertz and Weibull models are the limiting extreme value distributions for systems whose lifespan are determined of their first failed component \cite{GG01, Gumbel04}. 

Given the quantitative definition of aging in Eq. \ref{Eq One}, a system or an organism can be non-aging when $\mu(t)$ is a constant $C$, which indicates a constant chance of dying over time \cite{GG91, GG01}. 
In this kind of non-aging organisms, the drop of viability is exponential,  $S = e^{-Ct}$, and is identical to the exponential decay of radioactive isotopes. 
Intuitively, as long as non-aging individuals can live to the next day, their chances of survival will be as good as those on the previous day. 
In bacterial phages, drop of viability is  exponential  \cite{DePaepe06}, indicating their non-aging characteristics. 
(For clarity, non-aging is not immortal. It only means that chances of dying remain the same over time. ) 
It can be shown that constant mortality rate occurs when the Gompertz coefficient $G=0$, indicating that the rate of aging is zero in bacterial phages. 
It is therefore of importance to understand why aging occurs in many complex organisms but not in simple organisms.

Reliability theory is a well-established field in engineering \cite {BarlowProschan96,  Leemis09}, and its application in biological aging has been recognized decades ago \cite{Murphy78, Skurnick78MAD, Witten85MAD, Abernethy98JTB, Miller89JTB, GG91,GG01}.  
Murphy proposed a "Bingo model" in 1978 and considered an organism as a serial configuration of subsystems \cite{Murphy78} . 
Similarly, Skurnick and Kemeny in 1978 modeled an organism as a number of serial links, and recognized that the weakest link determines the organism's age \cite {Skurnick78MAD}. 
Witten in 1985 argued that organism can be modeled as a graph and explored ways to regenerate the Gompertz model using a serial configuration of components \cite {Witten85MAD}. 
Gavrilov and Gavrilova recognized the importance of non-aging components, and developed a sophisticated reliability model of aging \cite{GG91, GG01}. 
All of these previous reliability models are based on serially connected subsystems, analogous to serially connected fuse boxes, and lack many features of complex gene interaction networks  in biological organisms. 
Consequently, these previous models  have limited influences on molecular studies of aging. 

Cellular aging, also known as the Hayflick limit \cite{Hayflick61}, is the basis of physiological aging \cite{KirkwoodFinch02}. 
Molecular mechanisms of cellular aging are best understood in the budding yeast \textit{Sacchromyces cerevisiae}, a single-cell model model organism \cite {Schleit:2012:FEBS,  Blagosklonny:2009:Aging, Longo:2012:Cell-Metab,  McCormick12CG}. 
Aging of yeast cells can be measured by the replicative lifespan, the number of cell divisions that cells can accomplish before senescence, and the chronological lifespan, the duration of time that cells can retain their proliferative capability in the stationary phase \cite{Longo:2012:Cell-Metab}. 
Hundreds of genes were shown to influence aging when they were altered \cite{Kaeberlein:2005:Science, Powers:2006:Genes-Dev}, yet no gene has been shown to be a direct cause of yeast aging.
This perplexity is consistent with the general view that aging is a complex process and needs to be understood from the network perspective \cite{Managbanag:2008:PLoS-One, Promislow:2004:Proc-Biol-Sci, Smith:2007:Mech-Ageing-Dev, Soti07EG, Csermely06APB, Farkas2011Sci-Signal, Kirkwood97EXG, Kowald94JTB, Lorenz2009PNAS}. 
Moreover, genotypic contribution to cellular aging is estimated to be only around $20\sim30\%$ \cite{Qin06EXG}, indicating that cellular aging is a stochastic process to a large extent.
Despite the complexity  and stochasticity of aging, a universal characteristic exists at the population level: The log-transformed initial mortality rate and the Gompertz coefficient is negatively correlated, often termed the Strehler-Mildvan correlation \cite{StrehlerMildvan60, Qin06EXG}. 
This universality suggests a common principle underlying the stochastic aging processes in many organisms \cite{GG91, GG01}. 


To account for the complexity, stochasticity, and universality of cellular aging, we hypothesize that cellular aging is an emergent property of gene networks. 
An emergent property can be generally defined as a property found at the systems level but not at the component levels. To demonstrate the emergence of cellular aging from gene networks, we need to show that aging can occur in a gene network with only non-aging components by definition. 
Specifically, we will show that the key characteristic of biological aging, the exponential increase of mortality rate over time, can emerge from a gene network made up by components with only constant mortality rates. 
Our gene network model of cellular aging can not only provide fresh mechanistic insights on aging, but will also shed lights on network robustness and its connection to aging. 

\section*{Model and Results}

We will build our network model for cellular aging step-by-step. We will start with the simplest version - a single network module with static interactions, then build a modular network, and finally introduce stochasticity into the model network.  In this study, our main goal is to demonstrate the emergent aspect of aging at the systems level, and we will use only non-aging components in our model.

\subsection*{An essential network module with non-aging interactions}
\begin{figure}[!b]
  \begin{center}
    \includegraphics[width=4in]{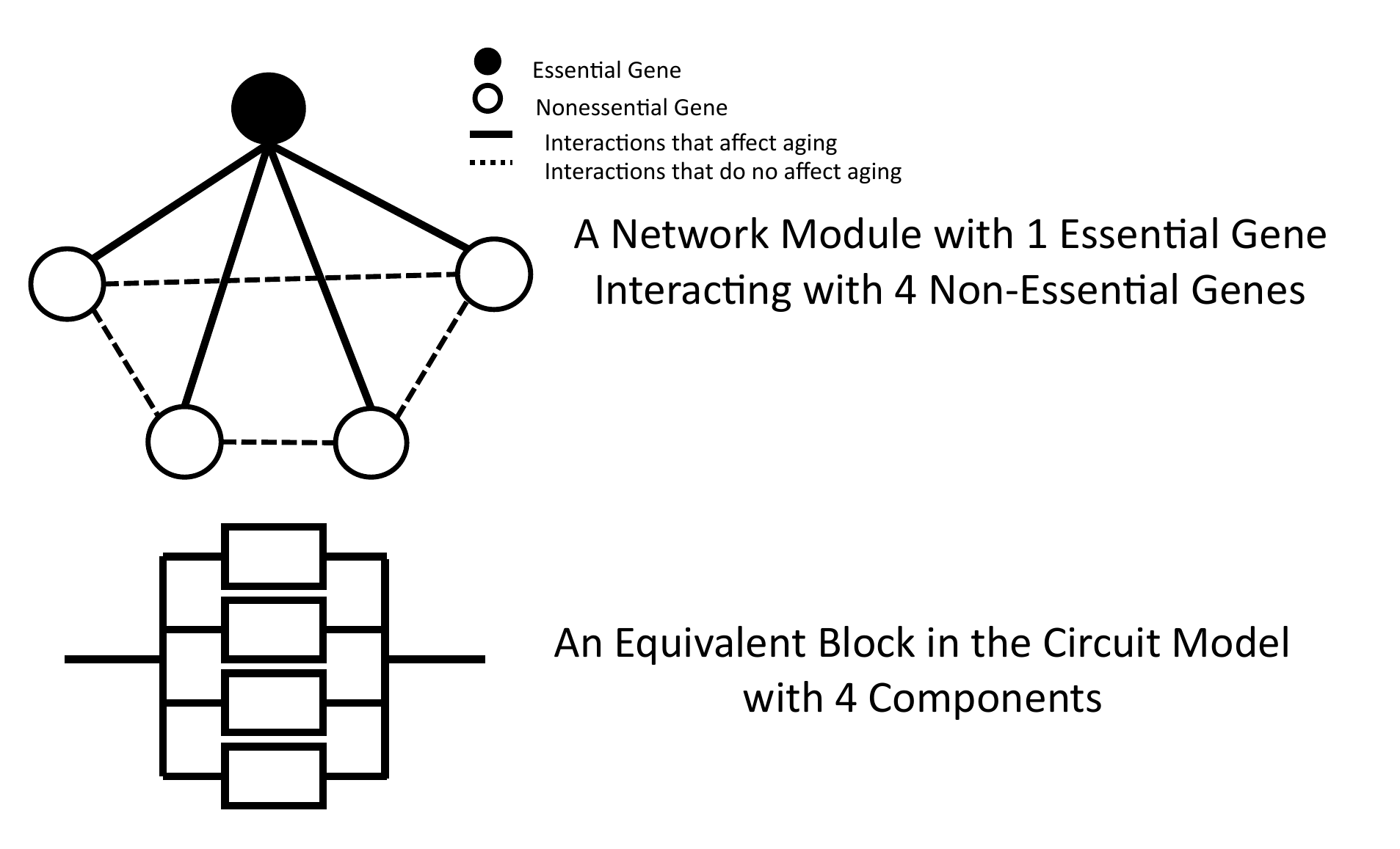}
  \end{center}
  \caption{\small A single essential network module and its equivalent block in the classical circuit model. In the circuit block, each parallel box is a component and can be viewed as a fuse. The circuit will die when all fuses burn out. 
 In the network module, circles represent genes: The solid black circle is an essential gene and open circles are non-essential genes.  Links between circles represent gene interactions.  
 Each interaction of the essential gene represents an biological activity that the gene performs.  The network module fails when the essential gene loses all of its interactions. 
Hence, only essential gene's interactions influence aging in the network model and are presented in solid lines. Each solid line in the network module is equivalent to a fuse box in the circuit model. 
 Dashed lines represent gene interactions among nonessential genes that do not influence aging, and they are not present in the classical circuit model. 
 Failures of gene interactions are assumed to be independent and follow the same constant decay rate. }
  \label{Single}
\end{figure}

We first introduce the concept of an essential network module - the basic building unit of our network model.  
There is one essential gene (solid black circle) and $n$ number of non-essential genes (open circles) in each essential network module (Fig. \ref{Single}). 
Each interaction represents a biological activity that a gene parcipates, and the decline of biological activity of each interaction is assumed to have the same constant mortality rate $\lambda$. 
For clarity,  declining activities of non-aging interactions will be termed 'decay', and the constant mortality rate will also be called the decay rate. 
An essential gene will cease to be active when it loses all of its interactions. 
This scenario is equivalent to the deletion of an essential gene, and leads to the failure of the entire network, i.e. cell death. 
The viability of each non-aging interaction is $e^{-\lambda t}$.  
To find out the viability of the module $s_m$ (the subscript $m$ represents module), we assume that decaying activities of gene interactions are independent. In other words, loss of one gene interaction will not affect the functions of the remaining gene interactions. 
This essential network module is equivalent to a circuit block with $n$ parallel components in the GG01 model \cite{GG01} (See Figure 5.3b in reference \cite {GG01}).
Failure of the essential module is analogous to the failure of a block in the GG01 model. 
Based on the reliability theory \cite{Leemis09, GG01},  the viability of the essential module is
\begin{equation}
 S_m(t) = 1 - (1-e^{-\lambda t})^n  \quad . 
\end{equation}
The mortality rate of the essential module, $\mu_m$ is
\begin{equation} \label {Eq mortalityrate.single.module}
 \mu_m(t) = - \frac{dS(t)}{S(t) dt} = \frac{n \lambda e^{-\lambda t} (1- e^{-\lambda t})^{n-1} }{1 - (1-e^{-\lambda t})^n} \quad . 
\end{equation}
When $t \ll 1/\lambda$, the above equation can be simplified to
\begin{equation} \label {Eq m.module.early}
 \mu_m(t) \thickapprox n \lambda^n t^{n-1},
\end{equation}
which indicates that in early life period, mortality rate of this essential network module increases as a power function of time. 

When $t \gg 1/\lambda$, Eq. \ref {Eq mortalityrate.single.module} can be simplified to
\begin{equation}
 \mu_m(t) \thickapprox \lambda,
\end{equation}
which indicates that mortality rate of the essential module drops to the value of a single component in late life period.

If there is only one interaction in the module, $n=1$, the module's morality rate is the constant $\lambda$. When the number of interaction $n$ increases, the module behaves like an aging system whose mortality rate, $\mu_m(t)$, increases over time.

\subsection*{A network with multiple essential modules}
\begin{figure}[!b]
  \begin{center}
     \includegraphics[width=3.5in]{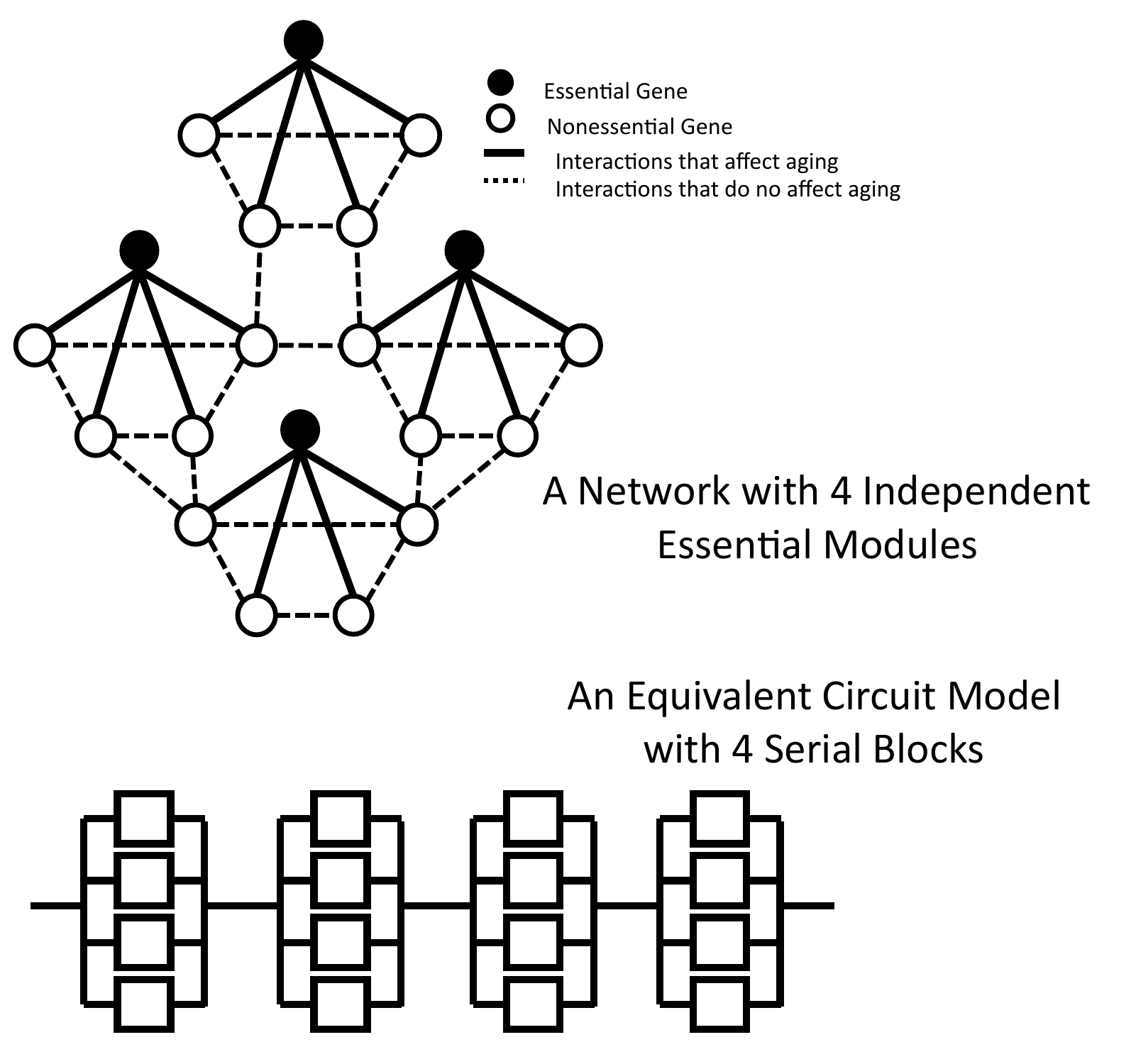}
  \end{center}
  \caption{\small A network with independent modules and its equivalent circuit model. Failures of essential modules are assumed to be independent, and interactions between essential genes are not allowed. Independent network modules are equivalent to the serially connected blocks in the circuit model. Interactions among non-essential genes do not influence aging, and their connection patterns can vary freely.}
  \label{Fig static_network_model}
\end{figure}

We can build a model gene network using the essential network modules. 
The yeast genome consists of about 6000 protein-coding genes with about 1000 of them are essential ones \cite {Winzeler99Science}. 
We can assemble $m$ number of essential modules to build a network model of aging as in Fig. \ref {Fig static_network_model}.  
We assume that failure of any essential module leads to the failure of the entire network and therefore cell death. 
This is a reasonable assumption because the absence of any single gene out of the 1000 or so essential genes leads to inviable yeast cells \cite {Winzeler99Science}. 
To obtain analytic approximations, we assume that essential genes do not interact with each other and their failures are independent. 
With these assumptions, the network model is mathematically equivalent to the serial construction of blocks in the GG01 circuit model.  
Based on the reliability theory, mortality rate of the entire system is the sum of the mortality rates of serial blocks \cite{GG91,GG01, Leemis09}. 
The mortality rate of the model gene network $\mu_{net}$ is thus the summation over the mortality rate of every essential module $\mu_{m,j}$ ($m$ is for module, and $j$ is the index for each module):
\begin{align}
 \mu_{net}(t) =                 &  \sum_{j=1}^{m} \mu_{m,j} = m \mu_m   & \label {Eq summation} \\
                 \thickapprox & m n \lambda^n t^{n-1}          & \text{ when $ t  \ll 1/ \lambda$},  \\
                 \thickapprox & m \lambda                            & \text{ when $ t \gg 1/ \lambda $}.
\end{align}
The summation step in Eq. \ref {Eq summation} is valid  because essential genes do not interact with other. 
This assumptions is only necessary to obtain analytic approximations here and can be relaxed in future simulation based studies. 

We can see that aging of the model network with multiple essential modules follows the Weibull model of aging in early life and plateaus in late life.  Our task now is to further modify this network model to produce the Gompertzian characteristic of biological aging -  the exponential increase of mortality rate over time.

\subsection*{Networks with stochastic interactions give rise to Gompertzian aging}
A key difference between machinery and organisms is the level of heterogeneity or noises in the systems \cite {GG91, GG01}. 
In machinery, components are often connected in the same patterns. 
In intracellular gene networks, gene interactions are inherently stochastic due to the limited number of gene products, noises in protein expressions, and the crowding nature of intracellular spaces \cite{Ghaemmaghami2003Nature, Newman2006Nature, Ellis2001Crowding}. 
Furthermore, transcription noises can be amplified into noises at protein levels \cite{zenklusen2008single}.
From a biological perspective, heterogeneity in gene networks indicates cell subpopulations with different properties, such as different sensitivities to stress factors.

\begin{figure}[!b]
  \begin{center}
     \includegraphics[width=3.5in]{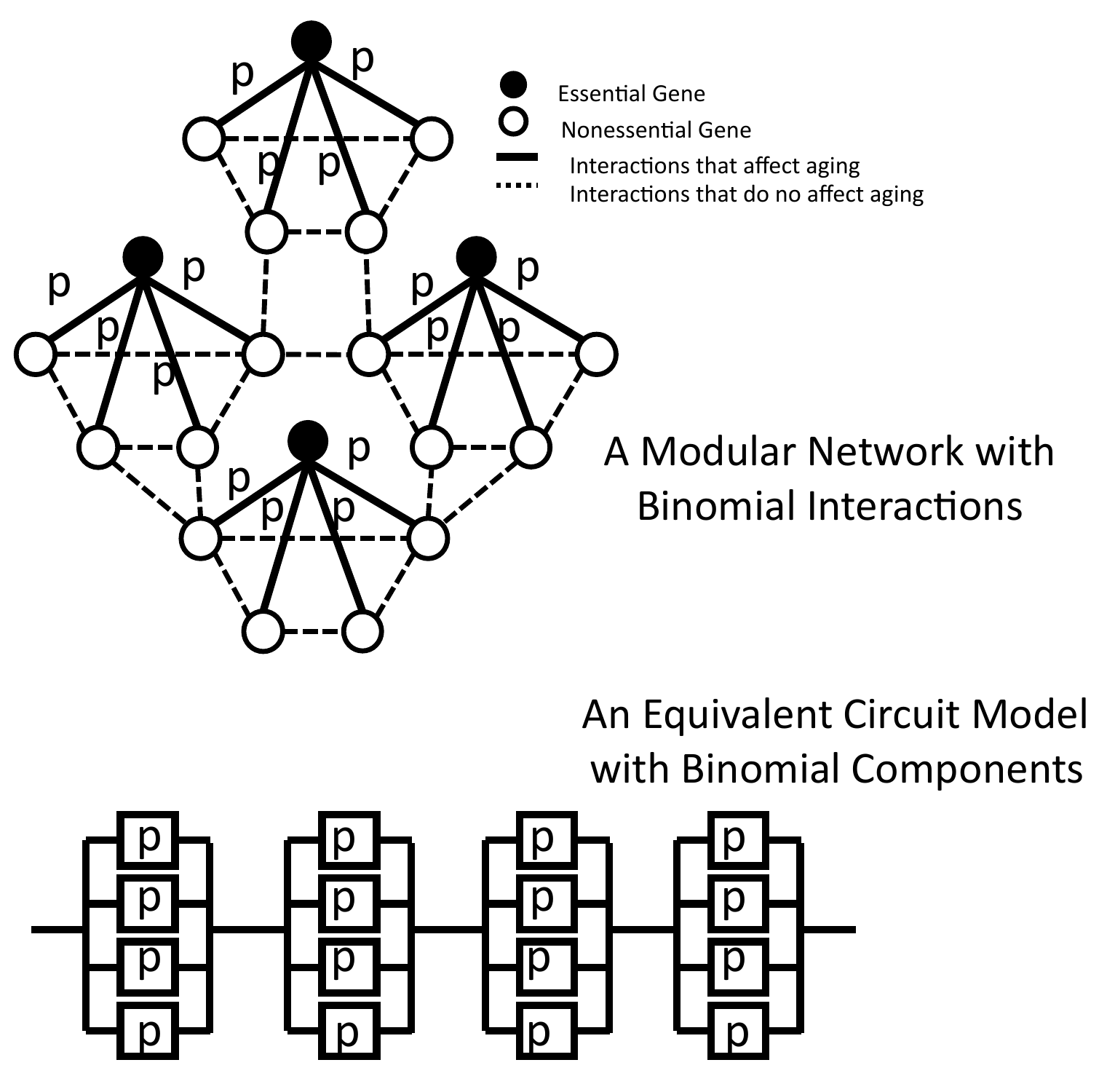}
  \end{center}
  \caption{\small A network model with stochastic interactions and its equivalent circuit model. In both models, the active number of components follows a binomial distribution, and the chance of a component being initially active is $p$.}
  \label{Fig stochastic_network_model}
\end{figure}

We will introduce stochasticity into gene interactions in our network model by assuming that the chance of an essential gene's interaction to be active is $p$ at time $t=0$ (Fig. \ref {Fig stochastic_network_model}). 
This binomial assumption changes the previous static network model in Fig. \ref{Fig static_network_model}  into a stochastic network model in Fig. \ref{Fig stochastic_network_model}. This stochastic network model is mathematically equivalent to the classic circuit model with binomially active components in GG01 \cite{GG01}. 
If a network that contains $m$ essential modules and each essential gene stochastically interacts with $n$ non-essential genes,  the mortality rate of the entire network is
\begin{align} 
\mu_{net} (t)	 \thickapprox & c m n \lambda p \sum_{i=1}^{n}\binom{n-1}{i-1} (p \lambda t)^{i-1} (1-p)^{(n-1)-(i-1)}  \quad , \label{Eq expanded binomial} \\
                                            & \quad \quad \quad \quad \quad \quad \quad \quad \quad \quad \quad  \text{when $t \ll 1/\lambda$ } , \nonumber
\end{align}
where $c$ is a normalizing constant, 
\begin{align}
c   =   \frac{1}{1-(1-p)^n}    \quad.  
\end{align}
It is reasonable to approximate the modular mortality rate as a summation of possible connection patterns in Eq. \ref{Eq expanded binomial} because we focus on the early stage of aging when $t \ll 1/\lambda$ \cite{GG01}. 
The summation term in Eq. \ref {Eq expanded binomial} is  the binomial formula $[(1-p) + p\lambda t]^{n-1}$, which leads to the following re-arrangements, 

\begin{align} 
  	\mu_{net} (t)  \thickapprox &    cmn (p\lambda)^n ( \frac{1-p}{p\lambda} +t  )^{n-1}  \label {n-1 binomial} \\
			 = &    cmn (p\lambda)^n (t_0 + t)^{n-1}  ,   \label{Eq binomial}
\end{align}
where 
\begin{align}
   t_0  =  \frac{1-p}{p \lambda} \quad.   \label{Eq t0}
\end{align}
The parameter $t_0$  has the unit of time and is termed the initial virtual age of the system, IVAS \cite {GG91, GG01}.
When $p=1$, we have IVAS=0, and the network mortality rate follows the Weibull model as previously described. When $p$ is non-zero, we will show network mortality rate can approximately follow the Gompertz model in the early stage of aging. 

We can re-write the network mortality rate in Eq. \ref{Eq binomial} using the IVAS, $t_0$, 
\begin{align} 
\mu_{net} (t) \thickapprox  cmn (p\lambda)^n t_0^{n-1} ( 1 + t / t_0)^{n-1}. \label{Eq IVAS} 
\end{align}
Gavrilov and Gavrilova recognized that the binomial term in Eq. \ref{Eq IVAS} can be approximated by an exponential form during the early stage of aging \cite {GG91, GG01}.  
A similar approximation was used by Witten for the viability function \cite {Witten85MAD}.  
Here, if we consider the range of time $t \ll t_0$, we can obtain the exponential increase of network mortality rate,
\begin{align} 
\mu_{net} (t) \thickapprox & cmn (p\lambda)^n t_0^{n-1} ( 1 + t / t_0)^{n-1} \nonumber \\
			 \thickapprox &  cmn (p\lambda)^n t_0^{n-1}  e^{\frac{n-1}{t_0}t} \nonumber \\
			 = & R e^ {Gt}       \quad \text{when $t \ll t_0$ }, \label {Eq Gompertz approximation}
\end{align}
where
\begin{align}
  R &=  cmn (p \lambda)^n t_0^{n-1}    =  mnp\lambda  \frac{(1-p)^{n-1}}{1-(1-p)^n}  \quad,    \label {Eq R} \\
   G &= \frac{n-1}{t_0}                        
   = \frac{\lambda (n-1)}{1/p-1} \quad . \label{Eq G}
\end{align}

The two conditions required for approximations in Eq. \ref{Eq expanded binomial} and Eq. \ref{Eq Gompertz approximation} are slightly different, $t \ll 1/\lambda$ versus $t \ll (1-p)/(p\lambda)$. 
These two conditions can be both readily satisfied, as discussed in GG01 \cite{GG01}.  When $p>0.5$, which is a biologically reasonable range (see simulations below), we will have $t_{0} < 1/\lambda$. 
In this case, $t \ll t_{0}$  always satisfies $t \ll 1/\lambda$. 
Intuitively, $1/\lambda$ is the average functional lifespan for gene interactions. 
Because the lifespan of the first failed essential module determines the lifespan for the entire network,  the average lifespan for gene interactions is much longer than the average network lifespan, especially for the yeast gene network that contains about 1000 essential genes. 
Consequently, when yeast cells die,  most of their network modules are still functional. 
Therefore, in the context of yeast aging, the two conditions for the approximations in Eq. \ref{Eq expanded binomial} and Eq. \ref{Eq Gompertz approximation} can be readily satisfied. 

Our results here show that the exponential increase of mortality rate over age, the defining characteristic of biological aging, can emerge from the proposed model gene network with only non-aging components.  
Therefore, we have demonstrated in principle that cellular aging is an emergent property of the proposed stochastic gene network model.

\subsection*{Variations in network configurations can cause the Strehler-Mildvan correlation}

The Strehler-Mildvan correlation indicates a negative relationship between the initial mortality rate $R$ and the rate of aging $G$, essentially arguing for a trade-off between the lifespan potential at birth and the dying-off phase late in life. 
We will show that this trade-off between $R$ and $G$ can be caused by changes of $n$, the average interactions per gene, and $p$, the chance of a gene interaction to be initially active, among populations. Because $n$ and $p$ can determine the network configuration, our results will argue that variation of network configuration can explain the the Strehler-Mildvan correlation. 

The influence of $n$ on the Strehler-Mildvan correlation can be shown by rearranging Eq. \ref{Eq R} and Eq. \ref{Eq G} to 
\begin{equation}
ln(R) = ln(M) - B G \quad,
\end{equation}
where
\begin{eqnarray} 
 M &=&  m n \lambda  \frac{p}{1-(1-p)^n} \quad,   \\
 B  &=&  \frac{1-p}{\lambda p} ln \frac{1}{1-p} \quad. 
\end{eqnarray}
Hence, when $n$ varies among different populations, it can lead to relatively large changes of G, but comparatively smaller changes in $ln(M)$, resulting in
a negative correlation between $G$ and  $ln(R)$, as argued by Gavrilov and Gavrilova  \cite {GG01}.   

\begin{figure}[!b]
  \begin{center}
     \includegraphics[width=3.5in]{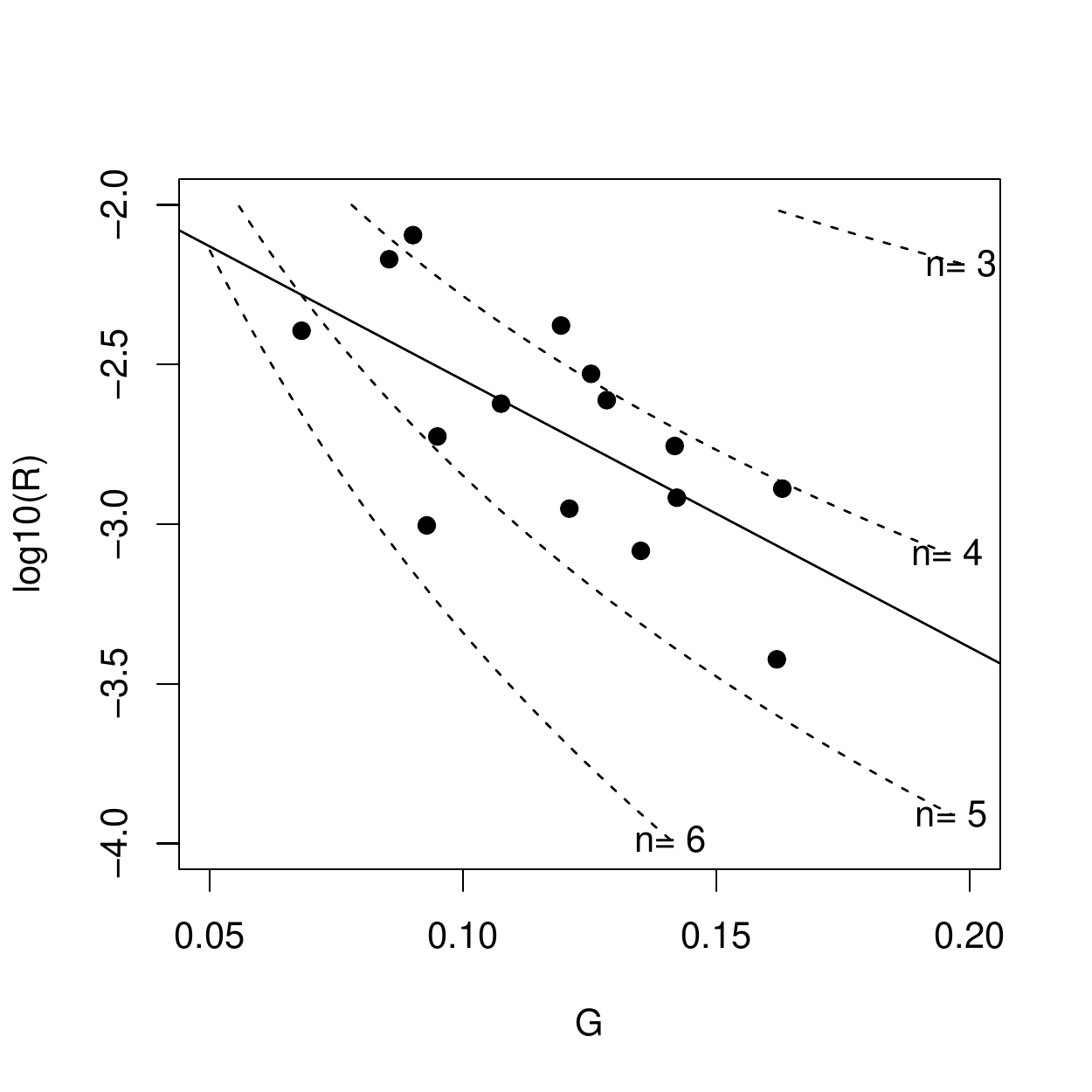}
  \end{center}
  \caption{\small Changes of network configurations can lead to the observed Strehler-Mildvan correlation in yeast natural isolates. The solid dark circles represent maximum likelihood estimations for natural isolates of yeast.  The solid black line is the linear regression model for the natural isolates, i.e., the Strehler-Mildvan correlation, with $p=0.0075$ and $R^2=0.44$ The  dashed lines are predicted results using our network reliability model with $\lambda=1/350$ and $m=1000$. The four dashed lines are for $n=3,4,5$, and $6$.  
  Each dashed line indicates the changes of $p$ when other parameters are held unchanged. The replicative lifespan measures include the data for 14 wild isolates published previously \cite{Qin06EXG} and additional data for the S288c strain measured in our group.    }
  \label{Fig natural isolates}
\end{figure}

The influence of $p$ on the Strehler-Mildvan correlation may not be obvious from Eq. \ref{Eq R} and \ref{Eq G}, but can be easily found by simulations, which we will discuss in the context of natural variations (see the dashed  lines in Fig. \ref{Fig natural isolates} discussed below). 

We will now show that changes of $n$ and $p$ can explain the observed Strehler-Mildvan correlation in yeast natural isolates. 
We found the Strehler-Mildvan correlation of $\log_{10}R = -1.72 - 8.4 G$ with a p-value of 0.0075 in a collection of yeast natural isolates \cite {Qin06EXG} (with additional data for strain S288c measured in our group). 
Based on the functional genomic data for the budding yeast  \cite {Winzeler99Science}, we can choose $m=1000$ for the number of essential genes.  
We can then  search  the parameter space for $n$, $p$, and $\lambda$, and look for parameter sets that can cover the observed $\log_{10} R$ and $G$ (see Fig. \ref {Fig natural isolates}). 
It can be seen that when $\lambda=1/350$, observations reasonably overlay with simulated values for $n=$ 4, 5 and 6. 
The influence of $p$ on $G$ and $log_{10}R$ can be seen in the dashed lines where $n$ is held constant. 
The trend of the dashed lines indicate that larger $p$ leads to smaller $log_{10}R$  and larger $G$ when $n$ is unchanged.  
To cover the empirical observations, the range of $p$ is about (0.9, 0.96) when $n=4$, and (0.83, 0.95) when $n=5$. 
Given that the yeast genome contains about 6000 genes with about 1000 essential genes \cite {Winzeler99Science}, these simulations show that variation of $p$ and $n$ in yeast populations is a plausible cause for the observed Strehler-Mildvan correlation.  

The above simulations are also helpful to understand IVAS $t_{0}=(1-p)/(p\lambda)$. 
To account for yeast natural variations, the range of $t_{0}$ is about (15, 40) when $n=4$,  and (20,72) when $n=5$. The average lifespan for wild yeast isolate is around 30 \cite{Qin06EXG}, which is in similar ranges of $t_0$. 
Hence, the period of $t \ll t_0 < 1/\lambda = 350$ exists for the Gompertzian characteristic of aging to emerge. 


In summary, the  Strehler-Mildvan correlation can be caused by variations in the average links per gene, represented by $n$, and the stochasticity of gene interactions, represented by $p$.   
Because the product $np$ represents the average number of active gene interactions that shapes the network configuration, population variation in gene interaction patterns may account for the ubiquity of this correlation. 

\section*{Discussion}
\subsection*{Implications of our network model of cellular aging}
Our work extends the classical serial circuit model into a modular gene network model. 
The inherent stochastic nature of gene interaction networks is  recognized as a key factor in the emergence of Gompertzian characteristic of aging. 
This simple gene network model can account for several important features of biological aging: the exponential increase of mortality over time \cite{Gompertz1825}, the ubiquitous Strehler-Mildvan correlation  \cite{StrehlerMildvan60, Qin06EXG},  
and the pleiotropic nature of aging \cite{Williams57}. 
The basic assumption of our model is consistent with the view that cellular aging is a process of stochastic accumulation of damages \cite{KirkwoodFinch02}.
Our current model essentially provides a parsimonious explanation for cellular aging. We show that it is sufficient to explain the emergent aspect of cellular aging using a simple stochastic gene network model. 

The importance of "the weakest link" in aging has been recognized many times \cite{Skurnick78MAD, Soti07EG, Farkas2011Sci-Signal}. 
Our simulation results in Fig. \ref{Fig natural isolates} offers a intuitive view on this effect. 
For a model gene network with $m=1000$ essential genes, failure of any essential module leads to cell death, which implies that most essential network modules are still functional when cells die. 
In our simulations with decay rate $\lambda = 1/350$, the average lifespan of a gene's activity is $1/\lambda=350$. 
The average lifespan for the entire network are mostly in the range of (20,50), only a fraction of the average component lifespan. 

How does the population-based model of aging as in Eq. \ref{Eq One} describe the aging process of each individual?
This is a frequently raised question. 
Basically, the aging process of each individual is modeled by the increasing chance of dying as in $\mu(t)$ and by the declining viability as in $S(t)$. 
From the viability $S(t)$, we can obtain the cumulative density function of lifespans as $1-S(t)$.  
Each individual lifespan is considered as an incidence drawn from the probability distribution described by $\mu(t)$ and $S(t)$.  
In other words, a defined model of mortality rate specifies the stochastic aging processes for individuals. 

\subsection*{Robustness and the rate of aging}
The role of network robustness in cellular aging can be appreciated by interpreting the non-aging feature in the death of bacterial phages from the network perspective. 
In the gene networks of bacterial phages, we predict that there should be negligible levels of functional redundancy or network robustness, and failure of any gene interaction will lead to system failures. 
Therefore, the mortality rate of the phage gene network  is the summation of the mortality rates of all of its components. 
Because the network mortality rate is constant , our model further suggests that biological activities in the bacterial phages most likely also decay in exponential fashion over time. 

In comparison, the emergence of aging in the yeast gene networks can be attributed to network robustness, which can be viewed as a generalization of redundancy in the classical serial circuit model of GG01  \cite{GG91,GG01}. 
Biological robustness can be defined as persistence of phenotypes in the presence of genetic, environmental, and stochastic variations \cite{Wagner2005Robustness}. 
Robustness of biological networks can be attributed to the overlapping roles of many biological activities or network buffering effect.
In the scope of cellular aging, robustness can be viewed as the maintenance of cellular homeostasis during the course of aging. 

Our model suggests that the Gompertz parameter $G$ is a good measure of network robustness. 
From Eq. \ref{Eq G}, it can be seen that the Gompertz coefficient $G$ is approximately correlated with $np$, the average number of active gene interactions per gene. 
It can be argued that more active gene interactions lead to more robust gene networks. 
Indeed, it often argued that aging is the most informative measure of biological robustness \cite{Soti07EG, Farkas2011Sci-Signal}  (also personal communication with Daniel Gottschling), a view that is consistent with the strong association between aging and human diseases. 

Counter-intuitively, the rate of aging, as defined by the Gompertz coefficient in Eq. \ref {Eq G}, increases when the active number of gene interactions increases (when $p$ or $n$ increases or both).  In other words, our model predicts that more robust gene networks have faster rate of aging. 
This prediction is counter-intuitive because we have partitioned the aging dynamics into the two Gompertz parameters. The rate of aging is described only by $G$ but not by $R$. 
In contrast,  the colloquial meaning of aging contains information for both parameters. 
The Gompertz coefficient $G$ reflects the transition phase in the survival curves. Larger $p$ and $n$ lead to increased $G$ but also decreased $R$, which will lead to extended plateau in the initial stage of life but more dramatic transition and a more sharper dying off phase in the survival curves.
An intuitive example is to compare the survival curves of normal population and those of centenarians. The transition of the dying off phase in the centenarians is much sharper, indicating higher $G$ value. 

\subsection* {Future directions}
The main focus of the current work is the theoretic plausibility of cellular aging as an emergent property of gene networks. Future work will emphasize on the application of this model to evaluate and interpret experimental data. 
Additional aspects of cellular aging will also be considered, such as renewals/repairs, network configurations, and complex gene interactions. 
In biology, new copies of gene/protein are often generated, old molecules are often removed, and some mis-functional molecules can be repaired. 
Gene networks tend to display power-law configuration and error-tolerant features. 
Currently, we considered only essential and non-essential genes. It is known that more complex scenarios of gene interactions exist in yeast, such as the synthetical lethal pairs.  
We do not think that these additional factors would change our main conclusion on the emergent aspect of cellular aging, but we do expect that these factors would likely improve the fitting of network models to experimental data. 

Cellular aging in yeast is often measured in genotypically homogeneous populations, and heterogeneity in these cells are non-genic. 
This kind of non-genic heterogeneity is modeled by the binomial distribution for active gene interactions in our network model. 
When parameters of binomial distribution vary among different yeast strains, it is clear that these parameter variations are due to genic variation, i.e. genetic diversity. 
Heterogeneity in populations is often modeled by frailty \cite{Vaupel1979Frailty}. 
Vaupel, Manton, and Stallard (1979) defined an "age-invariant" frailty as the ratio of an individual's mortality rate normalized by a reference individual's mortality rate \cite{Vaupel1979Frailty}. 
Baudisch (2011) proposed that the shape of survival curves can be measured by the ratio of the mortality rate of the long-lived individuals versus the initial mortal rate \cite {Baudisch11PaceShape}. The Baudisch shape measure is very similar to the frailty measure despite their differences in applications. 
It seems likely that the binomial distributions of gene interactions should be connected to the concept of frailty and its similar measures, but these connections remain as open questions. 

More generally, we think that gene interaction networks may be generalized as epistasis, i.e. genetic networks, and can then be applied for aging in multi-cellular organisms. 
It is known that mortality rate decreases over time in the tortoise \textit{Gopherus agassizii} and remains nearly constant in the freshwater polyp \textit{Hydra vulgaris} \cite{Baudisch12Sci}. 
It would be interesting to study how different factors can influence network aging and lead to these non-Gompertzian characteristics of aging.

\subsection*{Final remarks }

Our model demonstrates that aging can be viewed as an emergent property of gene networks and is connected to network robustness and stochasticity. 
Because gene networks are examples of biological complexity, we argue that in general, aging can be viewed as an emergent property of biological complexity. 
Our current network model offers a theoretic framework to explain the shared characteristics between replicative aging and chronological aging in yeast despite apparent differences in their involved gene pathways.  
We hope this work can persuade experimental biologists to examine cellular aging from a quantitative network perspective.  

\section*{Acknowledgments}

HQ thanks the support of the NSF Award 1022294, a summer support from the Undergraduate Faculty Research Residency program at the Fred Hutchison Cancer Research Center, a sabbatical fellowship from the National Institute for Mathematical and Biological Synthesis (NSF Award EF-0832858), and partial support from a HHMI award 52006314 and a NCMHD grant (NIH 5P20MD000215-05) given to 
the Spelman College.  HQ also thanks many friends and colleagues for fruitful discussions.  

\bibliography{qin_networkaging}

\newpage 
\section*{Figure Legends}

\end{document}